\documentclass[conference]{IEEEtran}
\usepackage [noadjust] {cite}
\usepackage{amsmath,amssymb,amsfonts}
\usepackage{algorithmic}
\usepackage{graphicx}
\usepackage{textcomp}
\usepackage{xcolor}

\begin{document}

\title{Deep Breath: A Machine Learning Browser Extension to Tackle Online Misinformation
}

\author{\IEEEauthorblockN{Marc Kydd}
\IEEEauthorblockA{School of Design and Informatics \\
Division of Cyber Security \\
Abertay University\\
Dundee, United Kingdom \\
m.kydd1800@abertay.ac.uk}
\and
\IEEEauthorblockN{Lynsay A. Shepherd}
\IEEEauthorblockA{School of Design and Informatics \\
Division of Cyber Security \\
Abertay University\\
Dundee, United Kingdom \\
lynsay.shepherd@abertay.ac.uk}
}

\maketitle

\begin{abstract}
Over the past decade, the media landscape has seen a radical shift. As more of the public stay informed of current events via online sources, competition has grown as outlets vie for attention. This competition has prompted some online outlets to publish sensationalist and alarmist content to grab readers' attention. Such practices may threaten democracy by distorting the truth and misleading readers about the nature of events. This paper proposes a novel system for detecting, processing, and warning users about misleading content online to combat the threats posed by misinformation. By training a machine learning model on an existing dataset of 32,000 clickbait news article headlines, the model predicts how sensationalist a headline is and then interfaces with a web browser extension which constructs a unique content warning notification based on existing design principles and incorporates the models' prediction. This research makes a novel contribution to machine learning and human-centred security with promising findings for future research. By warning users when they may be viewing misinformation, it is possible to prevent spontaneous reactions, helping users to take a deep breath and approach online media with a clear mind.
\end{abstract}



\label{introandbackground}
\section{Introduction and Background}

The Internet has rapidly become the dominant means for users to stay connected, even more so in the wake of the COVID-19 pandemic; however, this has led to several problems \cite{lallie2021cyber}. From connecting with friends and family, commenting on recent events, or just being entertained, the Internet plays an integral role in how we inform and stay informed \cite{mitchell_shearer_stocking_2021}. However, in a landscape that rewards attention rather than quality, some have turned to more dubious practices such as sensationalism, misinformation, and shocking imagery.

\subsection{The Incentivisation of Misinformation \& Clickbait}
As more and more content is published every second, focus shifts from producing quality content to hijacking attention. When users' interest is constantly pulled from one article, video, or image to the next, publishers need to find new ways to garner clicks. Consequently, this can easily lead to sensationalism, clickbait, and in the worst case, outright misinformation without the user being aware.

The constant news cycle of the Information Age presents difficulties for news outlets to write up each emerging story. Instead, some outlets adopt the practice of aggregating content, either lightly editorialising a piece of existing content or directing users to another source to read the information there. The method of news aggregation is controversial, with some feeling that it is blatant theft of content, whilst others see it as the only viable solution to surviving in a fast-paced information economy \cite{chyi2016parasite} \cite{coddington2020gathering}.

\subsection{The Impact of Sharing Misinformation and Clickbait}
Regardless of the ethics of clickbait and similar practices, publishing misinformation can be harmful once it reaches a broader audience. Many internet users utilise social networking, which compounds misinformation’s effects by allowing for rapid dissemination of falsehoods and half-truths. Often this creates a chain-reaction scenario where one user shares a story with another user, who in turn shares it with another, and so on. The impact of this phenomenon does not require a user to have a vast following on social media platforms. Côté and Darling \cite{cote2018scientists} found that once a Twitter account surpassed approximately 1000 followers, it was much more likely to attract other users from a wide variety of backgrounds, including members of the general public, outreach groups, and even influential decision makers. This dramatically increases the potential audience that a piece of content can reach - and be shared by. Such a situation can result in users being presented with the same information multiple times, albeit being posted by different users.

Fazio, Rand and Pennycook analysed the effects of widespread sharing of information \cite{fazio2019repetition}. The authors examined how repeated exposure to information (both true and false) affects individuals’ perception of the data presented. Participants were asked to rate a series of statements on a percentage scale based on their believability. Statements were presented in two ways: some were shown only once, while others were repeatedly shown and were interspersed with one-off statements. Participants’ perception of truth across the accuracy scale rose without any influencing factor, other than being repeatedly shown the same information. 

When rated by participants, information regarded as the truth was rated as such. Similarly, information strongly considered false did not influence participants sufficiently to cross over into the ``considered as true'' category. This was noticeable in the 1\%-30\% bin, where repeated exposure did not generate a pronounced change in participant opinion. One of the paper's key findings relates to information considered to be ambiguous. Information initially considered unclear gradually became regarded as accurate and authentic the more the user was exposed to the statement. This was particularly evident in the 41-70\% bin which saw, on average, an increase of 6 points compared to their original rating.

Existing research \cite{chyi2016parasite, coddington2020gathering, cote2018scientists, fazio2019repetition} suggests that, in the current hyper-social age, users risk having their perceptions warped by misleading statements, hyperbole, and repeated exposure to information. This repetition does not have to stem purely from repeatedly seeing the same news article displayed but also from the conversation surrounding the topic. Increasingly fragmented and confrontational viewpoints presented online complicates how users determine what is true or false. However, simply stating that a user is ‘misinformed’ or ‘wrong’ does not make them more inclined to reassess their understanding. Instead, a more thoughtful approach is required, giving users a starting point from which to come to their own conclusions. As Bronstein et al. \cite{bronstein2019belief} suggests, interventions catering towards the promotion of ‘open-minded and analytical thinking’ could be of benefit in curbing the impact of misinformation.

\subsection{Informing Users of Risk Online}
As misinformation has become more apparent and widespread in recent years, research into warning users when they are being misled has received increased interest, both in academia and industry. Many social media sites have taken steps to curb the impact of misinformation on their respective platforms, typically by presenting users with a warning label. This approach offers a simple means of quickly informing users that the content they are viewing may be misleading; proving popular with various social media platforms adopting similar practices including Twitter \cite{Roth2020} and Meta’s Facebook \cite{Meta2021}. 

Ross et al. \cite{ross2018fake} analysed the effectiveness of warning labels adopted by major social platforms. Focusing primarily on the methods used to dissuade users from sharing misinformation, the authors tested two different label styles, one replicating those used by Meta and another informed by contemporary research. Participants (N = 151) were shown content consisting of six true and six false stories. Group one were presented all stories without any warning label. Group two were shown half the stories with a warning and the other half without. The participants in each group were then asked to determine which stories were manipulated or fake and which were unaltered. The research found that neither of the warning messages changed user behaviour. Users did not appear to be more suspicious of labelled content and were just as likely to interact with the content as that which was unlabelled. 

\subsection{Designing Effective Warning Labels}
\label{warninglabels}
Ross et al. \cite{ross2018fake} indicate that providing additional context to the user can be beneficial in curbing the impact of misinformation. Careful consideration in the design process on what information is communicated to the user and how could be instrumental in limiting misinformation’s reach. 

Shepherd and Renaud \cite{shepherd2018design} conducted a literature review on designing effective security warning labels in browsers. Assessing existing work in this area, the authors found that time and resources are not adequately allocated to designing warning labels leading to user frustration and confusion. This sentiment appeared to be validated by Ross et al. \cite{ross2018fake}. The authors found that the effectiveness of current solutions differed. Some warnings wrongly assumed that users had background knowledge on a topic which decreased their effectiveness, whilst others were too vague in their language that users did not understand the ramifications of their choices. 

To combat the aforementioned issues, Shepherd and Renaud \cite{shepherd2018design} concluded with the proposal of a set of design guidelines for browser warnings. The authors note that warnings designed for privacy and those designed for security differed, suggesting that different priorities must be considered depending on the intended use- case. The proposed guidelines recommend using simple and concise language to alert users to a potential issue and using neutral colours to avoid an undue emotional response. Furthermore, the guidelines propose linking to additional resources should the initial description not prove sufficient. Although the research in question is primarily targeted toward warning labels for security purposes, there is still value in applying these recommendations to tackling misinformation. 

\subsection{Designing Effective Browser Warnings and Labels}
Much work has been done previously in the field of usable security concerning the design of warning labels. Early attempts at warning systems typically used contextual measures such as a small on-page popup informing the user of potential risk. However, work by Wu, Miller and Garfinkel \cite{wu2006security} illustrated these popups are often ignored, misunderstood, or users do not even recognise they are there. 

Further research in this area took on a different approach, utilising interstitial warnings. This approach required users to interact with the warning label before proceeding, the underlying theory of this approach being that making the warning the central focus of the users’ attention would increase the likelihood of users reading and making an informed decision on the contents of the warning. 

Until recently, most research on effective warning design has been limited to web-security topics such as expired certificates or phishing links. Kaiser et al. \cite{kaiser2021adapting} examined warning label design to inform users of potential disinformation online. Evaluating several different warning design styles, ranging from information-dense with minimal colouring, to warnings with a strong visual impact but minimal detail, the authors assessed how users responded to the designs in a realistic environment. 

In a survey conducted with 238 participants, the authors found that participants responded most favourably to designs which featured a reference to the perceived risk (“This page contains misinformation”) and the recommended next step (``Consider finding alternative sources.''). The authors note that none of the designs evaluated showed any significant difference in how likely users were to consult a second or alternative source afterwards. The authors propose that changes in behaviour were more likely to stem from the friction caused by having to manually click through a warning rather than the content of the warning itself. 

Multiple factors play a role in shaping how users respond to warning labels, including the language used within them. Findings from a research study conducted by Mozilla \cite{Mozilla2019} to understand how to design better warning labels highlighted that employing opinionated design was more important than providing objective information. This means it is more important to convey the idea of a threat rather than what the threat is - prior research suggested overly technical warnings lead to confusion among users. Mozilla implemented this by simplifying the warning heading to feature abstract but understandable language. Additionally, for scenarios where users want to know details of the underlying issue, the warning provides an accessible description of the risks associated with the security fault.

The issue of labelling misleading content is a challenging one. What counts as misinformation must be determined, and designing warning systems that promote critical thinking rather than knee-jerk reactions is still an ongoing area of research. Although the means of warning users adopted by major social platforms may have limited efficacy \cite{ross2018fake}, Shepherd and Renaud \cite{shepherd2018design} indicate that warning labels can cater to users’ assumed knowledge and understanding without provoking undue alarm or concern. Similarly, Kaiser et al. \cite{kaiser2021adapting} suggest that such research can be used to combat misinformation. 

\subsection{Detecting Clickbait with Machine Learning}
The vast array of content posted online every second makes it impossible for human moderators to assess and review all dubious uploads. The use of an automated system is merited, one capable of rapidly and reliably analysing content for potentially misleading information which can integrate intervention measures. Machine learning’s inherent capabilities for finding and predicting patterns in information are well-suited to tackling misinformation. Furthermore, machine learning has seen renewed interest over the past decade as computing power and data storage have matured to enable real-world applications across a host of use cases. 

Chen, Conroy and Rubin \cite{chen2015misleading} explored if clickbait, and by extension, misinformation, could be detected using machine learning methods. Conducting a holistic view of research in the field, the authors note four unique means of detecting clickbait. Initially focusing on the textual content of clickbait articles, the authors found clickbait often displays lexical and semantic features unique to its form. The authors cited work by Lex, Juffinger and Granitzer \cite{lex2010objectivity} which analysed clickbait based upon factors such as word length, word choice, and terminology, and found that a machine learning model could be trained to detect clickbait with 77\% accuracy regardless of the topic discussed. 

Appealing to users’ innate curiosity by using unresolved pronouns or alluding to content within the article was also consistent with clickbait. As a subsequent paper by Rubin et al. \cite{rubin2016fake} noted, automated fact-checking and verification systems could help detect language patterns in text and warn users that the content they are about to read may be misleading. The authors also note that such a tool could prove helpful for journalists too, alerting them when they may be conflating claims or accidentally misleading. 

Clickbait is not strictly limited to the text of the article in question; the authors also found that surrounding factors such as imagery and how the user interacts with the article play a key role. Regarding the former, the authors \cite{rubin2016fake}  cite Ecker et al. \cite{ecker2014effects} who found that clickbait articles were likely to feature images which were incongruent with the headline. In such articles, an image can be used to grab the would-be readers’ attention with an impactful but unrelated image or shape opinion before the article was read. 

The authors \cite{rubin2016fake} also noted that previous research had found clickbait outlets typically aimed to attract user attention before funnelling them towards sponsored content or advertising. Additionally, the time difference between “time spent reading the article” and “time spent sharing and commenting about the article” could also be a signifier of clickbait. In this aspect, clickbait articles tend to use alarmist or sensationalist headlines to provoke knee-jerk responses (whether that be commenting or sharing) before the user has actually read the content within.

\subsection{Problem Space}
The rapid rise of the information age has led some to adopt unethical practices to drive engagement. Whether these practices are deployed purposefully or not, they pose a serious risk to society. Although existing work has explored the use of warning labels, depending on how these are designed, these may be ineffective. Given the amount of content published every second, it would be impossible to label the accuracy of content manually. Instead, machine learning offers a compelling alternative. 
Misinformation poses a severe threat; therefore combining advances in machine learning and warning design means an effective solution can be proposed to keep users safe. 

\label{methodology}
\section{Methodology}
\begin{figure}[th]
\centerline{\includegraphics[width=0.45\textwidth]{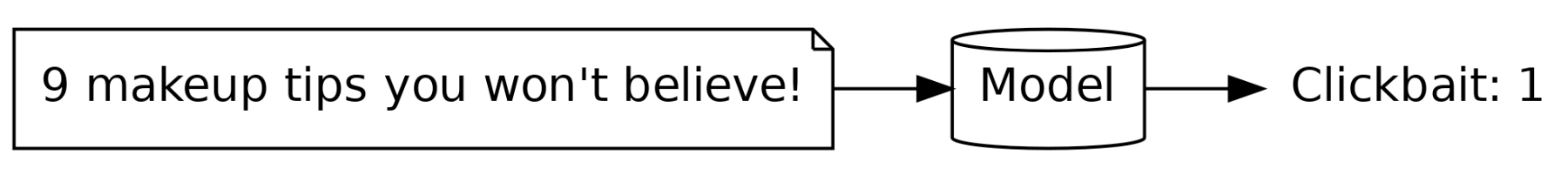}}
\caption{Example simplified data pipeline.}
\label{fig:Method_ML_Model}
\end{figure}

The proposed method consists of two-components: the machine learning model, for analysing and classifying content, and the web extension for communicating potential risk to the user. A simplified pipeline can been seen in Figure \ref{fig:Method_ML_Model}.

Using TensorFlow \cite{abadi2016tensorflow} and adopting the same dataset as used by Chakraborty et al. \cite{chakraborty2016stop}, a Sequential Model was trained on 32,000 news article headlines, labelled as either ‘clickbait’ or ‘non-clickbait’. The model, consisting of four layers (excluding the input layer), tokenises input text into a 64-dimensional dense vector before running it through a global average pooling filter. Output is then fed through a layer of ReLU nodes, followed by a final layer of Sigmoid nodes to arrive at a real number between zero (indicating neutral) and one (indicating strongly misleading).

The model was then connected to a browser extension via a Native Manifest, which allowed the browser to send portions of an article (e.g., the headline) to the model to analyse and generate a rating. The rating is then returned to the browser, after which a relevant warning can be presented to the user. Warnings were designed to be informative and actionable for the user, presenting clear detailing about the perceived risk and recommended next steps. 

\subsection{Developing the machine learning model}
\subsubsection{Dataset}
The same dataset used by Chakraborty et al. \cite{chakraborty2016stop} was adopted for this project and consisted of 32,000 news article headlines labelled as either ‘clickbait’ or ‘non-clickbait’. The dataset offered a robust and relevant base upon which to build. In particular, clickbait and misinformation rely on emotional language to provoke a response, suggesting the dataset would help develop a model well suited to detecting such language.

\subsubsection{Model}
In practice, the backend of this project centres around binary classification: Is this piece of text clickbait/misinformation or not? As such, using a Sequential model was deemed the most suitable due to its singular input-output structure, a structure well suited to classification tasks such as this.

\subsubsection{Preprocessing Data }
\begin{figure}[th]
\centerline{\includegraphics[width=0.45\textwidth]{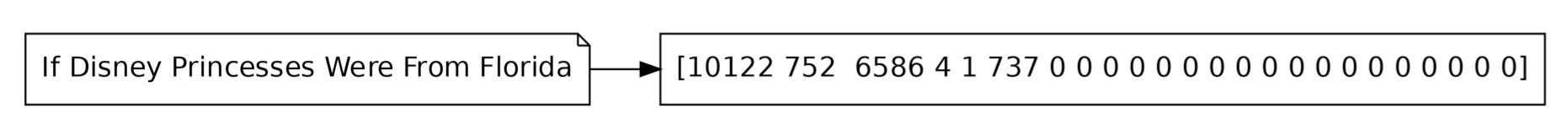}}
\caption{Example word embedding.}
\label{fig:Method_ML_Preprocessing}
\end{figure}

The dataset was split into 26,666 training samples and 5,334 testing samples which equates to ~83\% of the dataset for training and the remaining ~17\% for testing. Before training, the dataset had to be formatted such that the model could determine distinctions between clickbait and non-clickbait. This was achieved by converting, the headlines in the dataset into a vocabulary of word embeddings (Figure \ref{fig:Method_ML_Preprocessing}). To ensure uniformity across the dataset, all sequences were padded to 24 tokens long which was considered a safe maximum length for a headline. Headlines longer than 24 words long were automatically truncated to the maximum length. 

\subsubsection{Building the Model}

\begin{figure}[th]
\centerline{\includegraphics[width=0.45\textwidth]{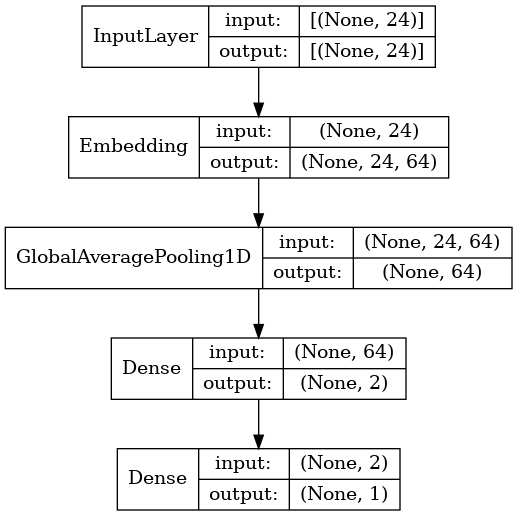}}
\caption{Visualisation of the Model.}
\label{fig:ModelViz}
\end{figure}

\begin{figure}[th]
\centerline{\includegraphics[width=0.15\textwidth]{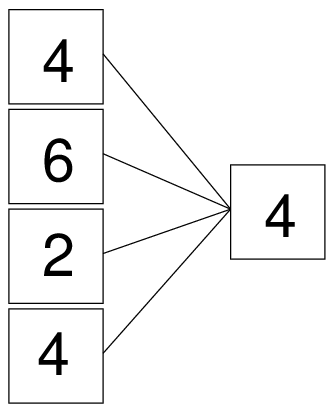}}
\caption{Visualisation of 1-Dimensional global average pooling.}
\label{1dpooling}
\end{figure}

The model is visualised in Figure \ref{fig:ModelViz}, and consists of the input layer, two hidden layers, and the output layer. The first layer takes the input (a tokenised sentence) and transforms it into a 64-dimensional dense vector. The usage of dense vectors allows for the semantic meaning of the sentence to be compressed, ensuring better generalisation. Despite their ability to derive underlying meanings and connections for a given sentence, the aforementioned dense vectors can result in overfitting if they become too detailed. To address the issue, the second layer consists of a Global Average Pool. This layer takes the 64-dimensional vector and determines the mean of each input channel (the 24-dimensional token sequences) which allows the model to learn approximations of embeddings rather than their exact values (Figure \ref{1dpooling}).

At this stage, the input data is now formatted and approximated to limit overfitting, and layers can be constructed, which will inform the output of the model. The first activation function of the model uses a rectified linear activation function (or ‘ReLU’). ReLU ensures that the next layer of the network receives a positive value as ReLU outputs 0 for input values equal to or less than 0 or the original value for those greater than 0. Finally, the data is passed through a Sigmoid activation layer, ensuring the resultant output falls between 0 and 1, i.e., “Is this piece of text clickbait or not?”.

\begin{figure}[th]
\centerline{\includegraphics[width=0.40\textwidth]{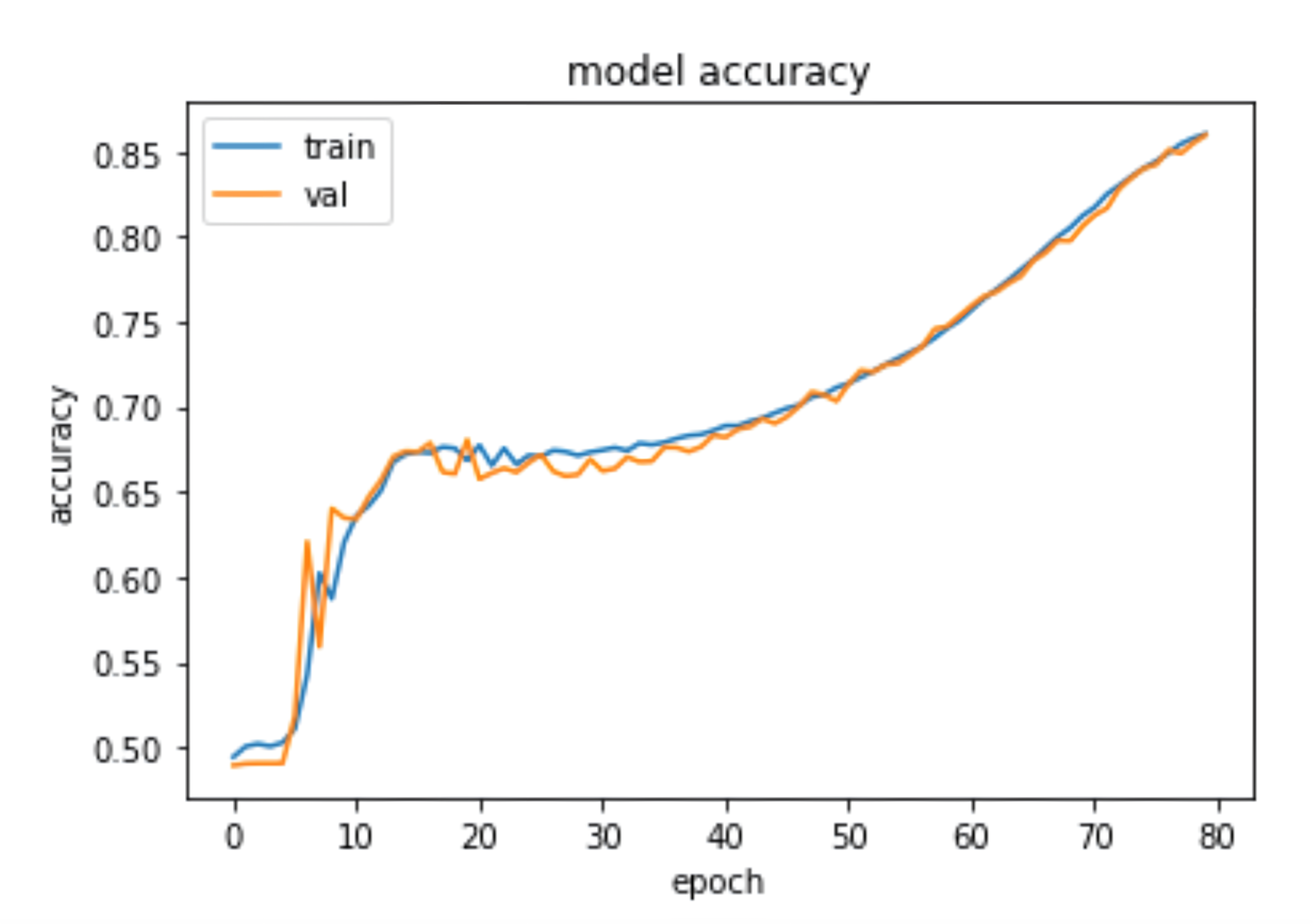}}
\caption{Accuracy graph during model training (Higher is better).}
\label{fig:Model_Accuracy}
\end{figure}

The model measures its performance based on the accuracy of its predictions. The task involves classification; thus, a binary cross-entropy loss function is used. The function calculates how far the models’ predictions stray from the dataset’s labels. A gradient descent with a momentum optimiser (also known as Stochastic Gradient Descent or SGD) further minimises the loss function. The optimiser helps improve the model’s training rate by minimising loss across training iterations. By doing so, it is intended that the model predictions will gradually trend towards the expected output.

\begin{figure}[th]
\centerline{\includegraphics[width=0.40\textwidth]{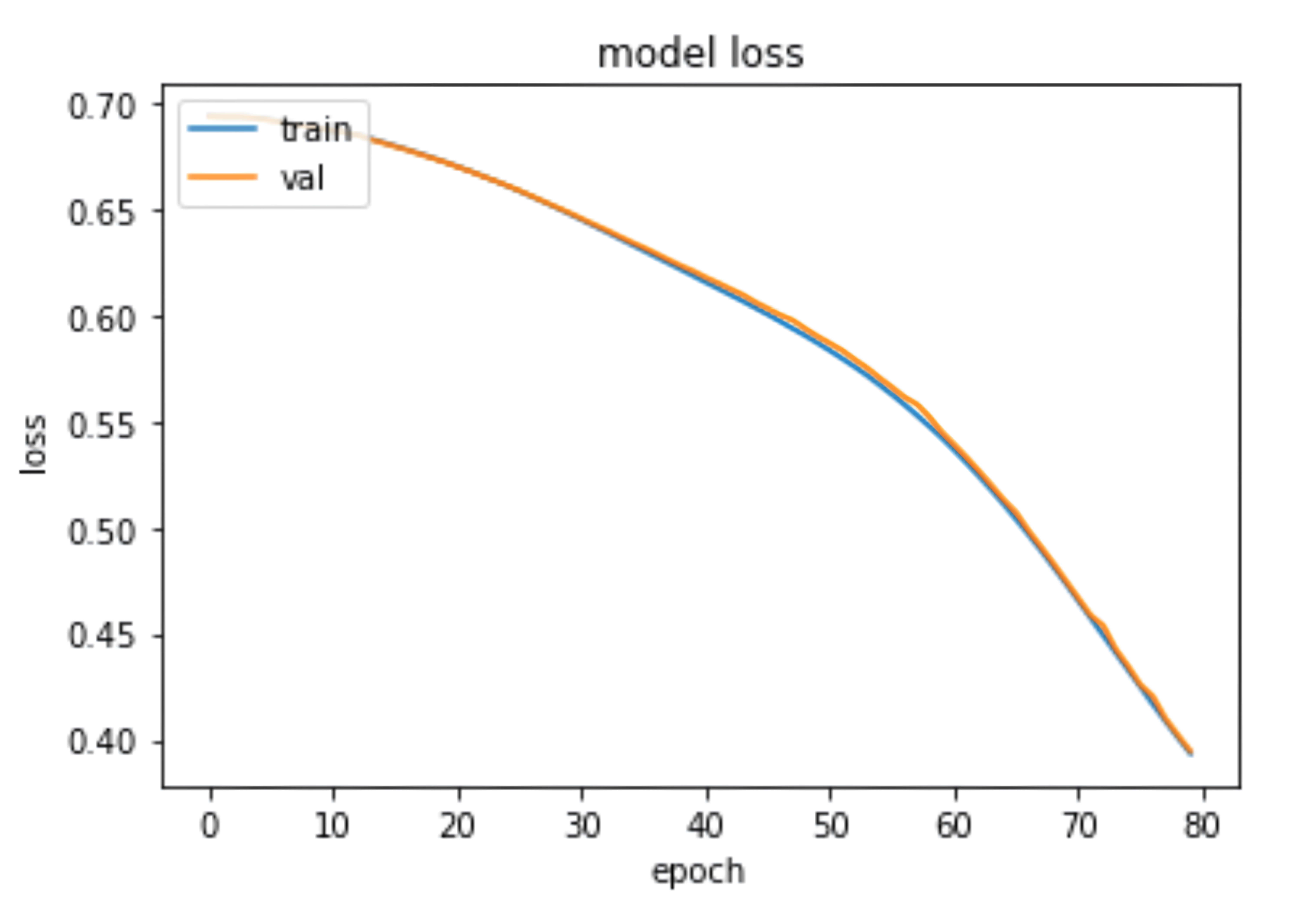}}
\caption{Loss graph during model training (Lower is better).}
\label{fig:Model_Loss}
\end{figure}

The model was trained for 80 epochs with a batch size of 128. Although batch sizes larger than 32 can lead to underfitting, it was intended that the larger epoch size would gradually result in greater accuracy and convergence, which can be seen in Figures \ref{fig:Model_Accuracy} and \ref{fig:Model_Loss}. After the model was compiled, trained, and evaluated, it was exported for use by the web extension.

\subsection{Web extension and warning messages}

\subsubsection{Creating the web extension}
A web browser extension was developed to ensure the model could be deployed in a real-world context. The extension means the model can analyse news articles as the user views them, providing a warning if misleading content is found. 

Native Manifests \cite{Mozilla2022} were used to allow a web browser to interface with a native application, passing data back and forth between the two. These manifests allowed the TensorFlow model to perform predictions locally on the machine and then send the resultant prediction to the browser for further analysis and output. 

Within the browser, the analysis begins when the headline of the page the user has visited is fetched. Initially, white space and control characters are trimmed. The headline is processed, and a value is returned indicating how sensationalist the headline is.

On the user’s device, the program transfers over to the native application, which handles parsing the headline into a format suitable for the model before computing a rating which is returned to the browser. Data from the browser is JSON-encapsulated and is sent via standard input (stdin), which the program reads from. Following this, the script begins importing libraries for loading the model and formatting the incoming data accordingly. The model is loaded, and the tokeniser is instantiated to convert the incoming headline. At this stage, the initial setup is complete and the model is ready for use.

The core of the script features a loop which waits for a message from the browser to be received, at which point the decoding process can take place. Now that the headline of the article is available, the script tokenises it and provides padding to ensure compatibility with the model. From here, a standard model prediction call can be made, encoded, and returned to the browser for display to the user. 

\subsubsection{Presenting warnings to the user}
When the native application produces a result, it is returned to the browser. The browser extension then generates a message sent to the news article's page, with the native applications result stored in a variable. This message is received by the content script, which is injected into each page by the extension.

The content script dynamically generates a warning label for the user. This is done by waiting to receive a message (the result) from the background script. This value is then multiplied by ten (to accommodate any floating-point issues that may arise (i.e. converting 0.8 to 8)). To prevent repeatedly warning the user about innocuous content, only headlines that score above five out of ten have a warning generated. Scrolling is disabled whilst the warning is on-screen, ensuring the user has to acknowledge it.

With regards to this project, the existing literature points towards interstitial warnings being the most likely to promote change in user behaviour. Additionally, even if most users do not appear to actually read the content of a warning label, they do show a preference for such information being present.

Informed by the papers discussed in Section \ref{warninglabels}, the web extension was designed to ensure strong visual clarity to effectively convey risks associated with a piece of misleading content. 

\begin{figure}[th]
\centerline{\includegraphics[width=0.45\textwidth]{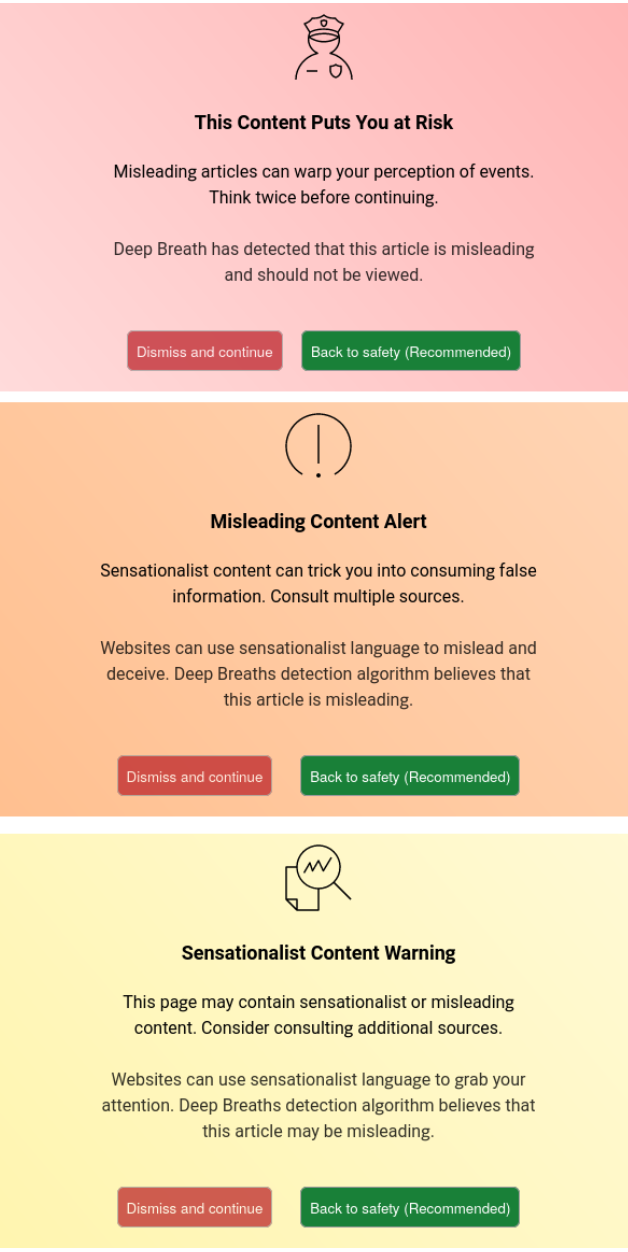}}
\caption{Sample of warning designs.}
\label{fig:ExemplarWarnings}
\end{figure}

The warning adopts a paywall-style design, mimicking an approach that users will likely already be familiar with from other news sites. This helps to ensure that the warning is not overlooked, which can happen with contextual warnings. To further ensure that the warning is brought to the user's attention, an overlay is used to darken the article and scrolling is prevented while the warning is on screen. The warning design comes in 5 variants - ranging from most minor to most severe, depending on the article's rating. Exemplar designs can be seen in Figure \ref{fig:ExemplarWarnings}

A vital problem with previous warning designs is the poor communication of risk, whereby warnings may be obscured by jargon \cite{kaiser2021adapting}. The design of the warnings seeks to minimise existing issues by conveying as much relevant information as possible in an easy-to-read format. Prominent symbols represent increasing risk levels based on the article's rating. Articles lower on the risk scale are given a more general 'magnifying glass' symbol, promoting the notion of thinking more critically about the article's merit. If an article includes more severe levels of misinformation, increasingly prominent 'alert’-oriented symbols are deployed, such as warning signs, stop signs and symbols of authority such as police figures. 

Additionally, an oscillating gradient is placed behind the warning. Depending on the severity of the warning, the colour used will shift from yellow to orange to red. The subtle movement of the gradient is intended to draw the user's eye to the warning, with the unique colour of each warning also helping the user understand the associated level of risk. 

Ultimately, the extension seeks to change user behaviour and provide education on meaningful steps users can take to protect themselves from misinformation in the future. As such, each warning label features unique phrasing that informs the user of not just what the perceived risk is but also advice on actionable next steps. 

Two buttons are presented to the user at the bottom of the warning, allowing them to dismiss the warning and continue, or navigate away from the page. To indicate the intended behaviour, the option to navigate away is displayed in prominent green with a 'Recommended' label included in brackets. Conversely, the option to dismiss the warning is presented in red and is slightly faded out to deliberately be obscured against the background until the user hovers over the button. 

\label{results}
\section{Results and Discussion}
\begin{figure}[th]
\centerline{\includegraphics[width=0.45\textwidth]{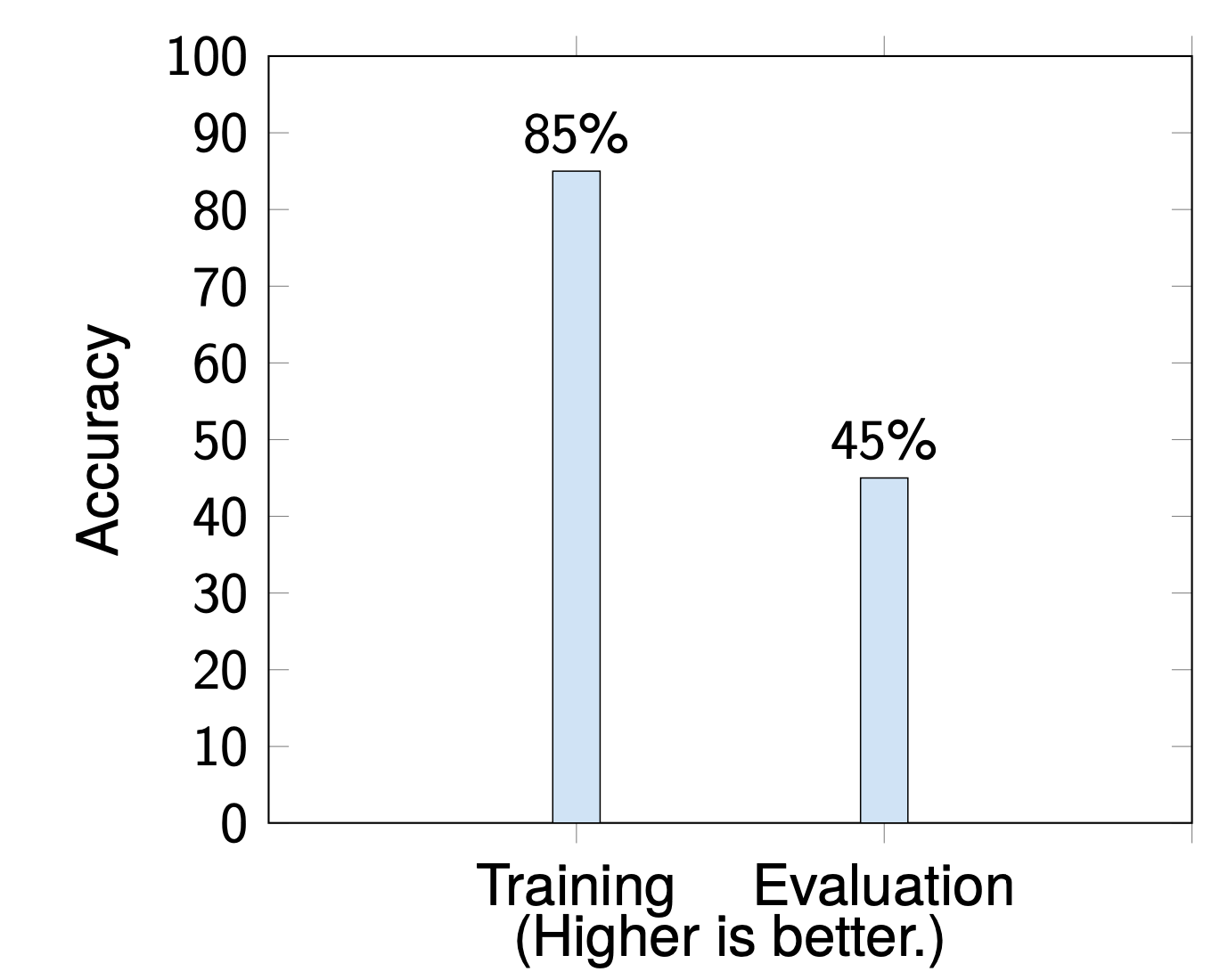}}
\caption{Accuracy comparison between training and evaluation.}
\label{fig:ML_Training_Accuracy}
\end{figure}

\begin{figure}[th]
\centerline{\includegraphics[width=0.45\textwidth]{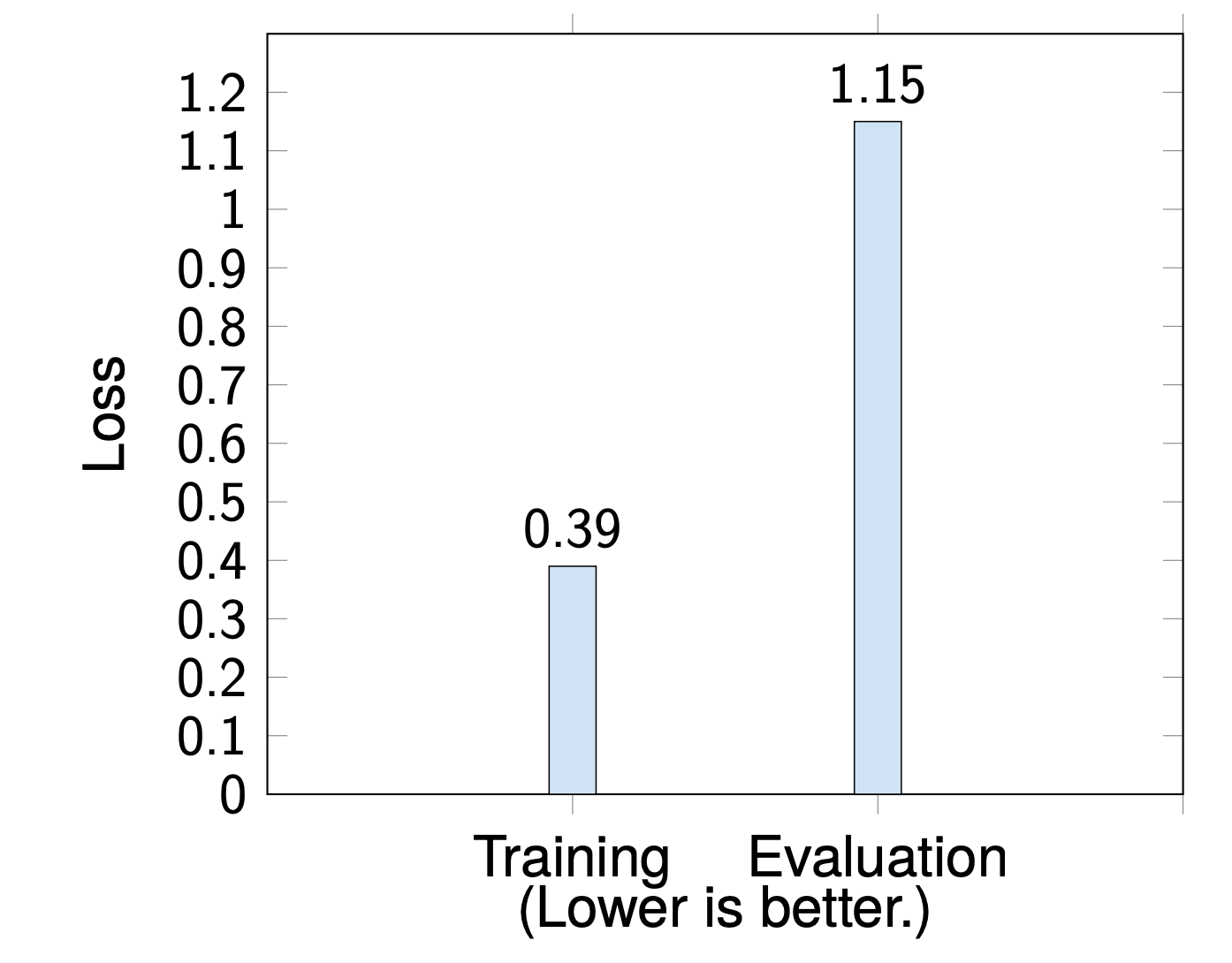}}
\caption{Loss comparison between training and evaluation.}
\label{fig:ML_Loss_Accuracy}
\end{figure}

During training, the model achieved an accuracy of approximately 85\% and a loss of 0.39, and when pitted against the evaluation dataset, the model achieved an accuracy of approximately 45\% with a loss of 1.15 (Figure \ref{fig:ML_Training_Accuracy}, Figure \ref{fig:ML_Loss_Accuracy}). This decline in accuracy likely stems from inconsistencies in the existing evaluation dataset, e.g., improperly formatted data, such as some of the labels assigned to the headlines do not appear to be correct. This could be resolved via an additional data cleaning. Another point of note is that the evaluation dataset used only binary labels (Is this headline ‘clickbait’ or ‘news’?), which may have also contributed to the discrepancy in accuracy as the model was producing a result between 0 and 1 instead of a pure binary output. 

With regards to machine learning, the project confirms the findings of Lex, Juffinger and Granitzer (2010) that clickbait and misinformation can be detected based upon lexical semantics, namely word choice, word length, and word commonality, i.e., Word \textit{x} appears frequently alongside word \textit{y}.

\label{conclusion}
\section{Conclusion and Future Work}
The model demonstrated in this paper has shown a reliable degree of performance, however, it could be refined further to derive even better results. The models’ accuracy and loss were still increasing and decreasing, respectively, suggesting better performance could be obtained before the curves flattened out. Furthermore, the capability of the model could be extended further. The model has been trained only on clickbait-styled headlines, which was effective. However, more robust results may be achieved by training on the contents of clickbait articles which would allow the model to develop a deeper understanding of the article and make a more nuanced prediction. The model used in this paper is a Sequential model designed to take a single output and produce a single result. Although this is effective at classifying a single headline as used in this paper, greater functionality could be achieved by allowing multiple inputs and outputs. This could include assessing the article’s headline but also a selection of sentences from the article. 
The rating assigned to content is dynamic; however, the underlying warning remains static. By expanding the models’ capabilities, it may also be possible to provide personalised warnings relevant to the content. In practice, this could mean warning the user about specific aspects of the article, such as sensationalist authors, misleading sentences, and miscaptioned images. Although every effort was taken to ensure the model made balanced and accurate predictions, no system is infallible. Conducting user testing and introducing the option for users to report when the model makes a perceived miscalculation could help adjust for missteps. 

The work presented in this paper makes promising advances toward tackling the issue of misinformation online by combining machine learning, human-computer interaction research, and web technologies. Findings validate and build upon prior research, and incorporating machine learning with usable security is still a relatively under-explored area of study.

\bibliographystyle{IEEEtran}
\bibliography{references.bib}

\end{document}